# THE DILEMMA BETWEEN EUPHORIA AND FREEDOM IN RECOMMENDATION ALGORITHMS

James Brusseau

Abstract: Today's AI recommendation algorithms produce a human dilemma between euphoria and freedom. To elaborate, four ways that recommenders reshape experience are delineated. First, the human experience of convenience is tuned to euphoric perfection. Second, a kind of personal authenticity becomes capturable with algorithms and data. Third, a conception of human freedom emerges, one that promotes unfamiliar interests for users instead of satisfying those that already exist. Finally, a new human dilemma is posed between two types of personal identity. On one side, there are recommendation algorithms that locate a user's core preferences, and then reinforce that identity with options designed to resemble those that have already proved satisfying. The result is an algorithmic production of euphoria and authenticity. On the other side, there are recommenders that provoke unfamiliar interests and curiosities. These proposals deny the existence of an authentic self and instead promote new preferences and experiences. The result is a human freedom of new personal identity.

Gli attuali algoritmi di raccomandazione dell'intelligenza artificiale generano un dilemma umano tra euforia e libertà. Per chiarire tale dinamica, si delineano quattro modalità attraverso cui i sistemi di raccomandazione trasformano l'esperienza. In primo luogo, l'esperienza umana della comodità viene modulata fino a una perfezione euforica. In secondo luogo, una forma di autenticità personale diventa catturabile mediante algoritmi e dati. In terzo luogo, emerge una concezione della libertà umana che non si limita a soddisfare interessi già esistenti, ma promuove interessi ancora inesplorati da parte dell'utente. Infine, si configura un nuovo dilemma identitario tra due modalità di costruzione del sé. Da un lato, vi sono algoritmi che individuano le preferenze centrali dell'utente e le rafforzano offrendo opzioni simili a quelle già risultate gratificanti: ciò genera una produzione algoritmica di euforia e autenticità. Dall'altro lato, vi sono agenti di raccomandazione che sollecitano curiosità e interessi inediti: tali proposte negano l'esistenza di un sé autentico, promuovendo invece l'emergere di nuove preferenze e di nuove esperienze. Il risultato è una forma di libertà fondata sulla possibilità di una rinnovata identità personale.

Recommendation algorithms are everywhere. They determine the songs we hear on Spotify, the films we watch on Netflix, the occupations we find on LinkedIn, the social circles we join on X, and who we will love — or, try to love — on dating platforms like Tinder. As more aspects of our lives move online, so too more of our experiences fall under the control of artificial intelligence suggestions.

A continuing longitudinal study of couples in the United States reveals that almost none met through online platforms in 1990, but today the percentage has climbed above 60, more than every other way of meeting combined. So, it is not just that increasing aspects of our human experiences are being touched by algorithmic filtering, it is that within each aspect the influence is expanding.

Once, human experience included recommendation algorithms, now the recommenders shape human experience.

The purpose of this paper is to begin mapping this reshaping, and to show how recommendation algorithms are changing the kinds of experiences we can have.

Specifically, four clusters of questions will be pursued. One concerns convenience. What potential for convenience emerges as recommendation technology advances? What does perfect convenience mean, and what will we sacrifice for the automated ability to get what we want? The second questions concern human authenticity. How much can predictive algorithms tell us about who we are as unique persons? What does uniqueness mean in this context? The third group of questions address human freedom. What kinds of human autonomy are enabled by recommenders? Which kinds are created? Finally, questions will be asked about personal identity. Today's recommendation algorithms are built on data about who we have already been. What does this mean for who we could be, and how do recommenders intersect with the human potential to change, to become someone else?

## 1. Euphoria convenience

On two levels, recommendation algorithms produce user convenience. Prosaically, they streamline discovery. As digital curators, they make it easier to find engaging films on Netflix, professional opportunities on LinkedIn, matches on dating sites (Abbas *et al.* 2022).

Above this prosaic convenience, there is one more radical. It is not just that we get what we want faster, it is that we get it before we know that we want it. The logical endpoint of recommendation strategies is anticipation. After reducing the multitude of possibilities to a manageable set, and then after reducing that set to a single best choice, the remaining challenge is to offer that choice at the critical moment: just before it is sought. This convenience is so perfect, in other words, that there is no time for inconvenience. There is no time to ask what we want or how to get it.

We already have hints of this convenience when Netflix begins running a recommended movie on one side of the screen, just as the last movie ends on the screen's other side. The idea is that the viewing experience continues seamlessly. The question "What should we watch now," or, "What should we do now," never comes up. The satisfying answer is given before anything can be asked.

How much of our experience will be exposed to this euphoria of convenience remains an open question, one to be decided by advances in data gathering and processing to predict the oncoming psychological states of users (Yang *et al.* 2024). What is certain, though, is that wherever human behavioral cues can anticipate interests or needs, there is potential for algorithmic convenience that is not less than perfect, but *more* than perfect. The problem — if there is one — is not that the recommender is insufficiently good, but that it is too good.

It is so good that it breaks convenience's conceptual limit: convenience peaks as its own disappearance. Getting what we want before we know that we want it directly means the end of inconvenience, but it also means that the dialectical idea of convenience can find no meaning: because one without the other is nothing, the entire question of convenience and inconvenience evaporates.

More, personal freedom disappears at the same instant. The act of freely choosing becomes redundant because getting what we want comes most easily by not making any choice at all. Stronger, choosing is a contradiction since any choice can only be inferior to what we already have, and so defeat the reason for making a selection in the first place. In the end, this convenience of perfect satisfaction is a state where having the freedom to choose is literally an inconvenience.

## 2. Authenticity

Euphoria convenience is also, potentially, a termination of the search for personal authenticity. For Martin Heidegger (1927), the idea of authenticity meant aligning our lives with our own unique purposes and projects. His ethical compass was not steered by external duties, it was not oriented by the metaphysical Good as Plato proposed, and it was not aligned with perfected rationality as Descartes and Kant advocated. Instead, authenticity was always internal and accessible. For each of us, authenticity meant discerning our own singular identity, and incarnating it as distinct from all those around us. To say, finally, that Heidegger's authenticity meant being true to oneself would be a gross simplification, but it would not be inaccurate.

The difficulty in Heidegger's philosophy is knowing ourselves. In a crowded world governed by social conventions and massified behaviors, it can be challenging for individuals to discern their own unique ambitions, fears, and hopes. To do that, Heidegger's instructions were to face our own death, to take it seriously. From that foundation of singularity and finitude — from the realization that no one can die for me — there comes a resoluteness. This is an ability to take responsibility for my own life, and to soberly determine what to do with the time that remains. This determining is the realization of authenticity. It is how I learn who I am, uniquely.

For euphoria convenience, the process of authenticity is very different. It is less human and more mathematical. Instead of existential experiments there are algorithms. Instead of resolutely facing our own deaths, there is data organization. The authenticity question as posed by euphoria convenience is: can our true and identifying needs, fears, and desires be assembled for us by binary information machines? Can our authentic selves be discerned by filtering data about our lives? Can, finally, algorithms construct a picture of individual users that is as true (or truer) than the one that users can self-discover?

About data and recommenders, Apple CEO Tim Cook has said:

> Scraps of data, each one harmless enough on its own, are carefully assembled, synthesized, traded, and sold. This process creates an enduring digital profile and lets companies know you better than you may know yourself. (Evans 2018)

What is being contemplated here is the possibility that, just as calculators perfect our human ability to multiply and divide, recommendation algorithms could perfect our self–knowledge. More reliably than our inner–reflections, they could determine the occupations we are meant to fill, the places we are most organically suited to live, the women and men who will truly be our most compatible and supporting partners. And, they could do all that as reliably as calculators do math.

Of course, no one believe that today's recommendation algorithms can circumscribe users, and few believe that inert algorithms and data will ever be able to delineate human being, but it nevertheless remains true that authenticity is on the horizon of recommendation technology.

## 3. Freedom to choose

While recommendation algorithms numb human freedom with euphoria convenience, they can also vitalize freedom on two levels: choice and creation.

Choice freedom is the ability to make purposeful selections within the context of overabundant options. This is not the ability to choose better or worse possibilities but, before that, the ability to make any meaningful selection.

The paradox of abundant choices overwhelming the human ability to make any intentional selection was rehearsed by Jorge Borges (1941) in his short story *Library of Babel*, which featured a library holding every alphabetically possible book, every combination of letters, spaces, and punctuation was written out and gathered as a volume on the endless shelves. This interminable book collection presented the tantalizing problem. The perfect novel and the ideal poem were in there, but actually *finding* them was practically impossible. Randomly pulling one or another volume from the countless stacks was futile, a mockery of the idea of choosing.

An analogous problem faces today's subscribers to Spotify and Tinder. The irresistible song, the perfect soulmate are literally there in your hand. But, distinguishing them from all the others in the electronic library is an initially hopeless task. Spotify (2024) has a catalog of 100 million songs, Tinder has 50 million users ("Wall Street Journal" 2024). With so many, there are no options at all. There is no intentional choosing until a provisional filtering reduces the random chaos to a manageable set (Ricci *et al*. 2022). When recommenders do that filtering, they are not just helping to make a choice, they are making meaningful choosing possible. They are converting a state of elective despair into an intentional reality where distinctions can be capably drawn, and where personal preferences can be expressed.

## 4. Freedom to choose differently

Besides making choosing itself possible, recommenders promote human freedom creatively, they open horizons of unfamiliar interests and experiences. Here, freedom is not so much about being able to make particular selections, but about opening new ranges of potential selecting.

When a restaurant recommender proposes a location that is unfamiliar not only because it has not yet been patronized, but also because of the style of food it serves, then it is proposing a new *kind* of place.

And when that happens it is not just a single restaurant that opens for the user, but a whole set of subsequent possibilities not previously envisioned. So, a pasta lover may be pleased to find that an AI recommender has proposed a previously unknown Italian restaurant, but it is a more creative possibility when an Asian–Italian fusion restaurant is suggested because dining there may open the way to a deeper interest in Asian cuisine, and then to an entirely unexpected preference and subsequently a host of new restaurant possibilities. A whole set of Asian restaurants that would never have been entertained are now spun into the recommendation funnel because of that single, original and creative algorithmic proposal.

Analogously, an online learning platform that had been responding to a user's established interest in conventional, representational art may recommend a lecture on Pablo Picasso's *Demoiselles d'Avignon* (1907), and that work may be described as a precursor to cubism. Then, that recommended subject may lead to another lecture on Georges Braque's *Houses at L'Estaque* (1908), which is sometimes understood as paradigmatic cubism. The extended result may be a lengthening set of learning opportunities stretching through the broad Cubist artistic movement. Technically, this passage would include discussions of geometric shapes, and the origins of influence, and the abandonment of traditional perspective in painting, but in term of human freedom, the movement of recommendations would be radically creative. A new horizon of art appreciation and interest opens. Because of the freedom recommendation, innumerable subsequent paintings become visible to explore.

Of course, these kinds of serendipitous moments are common in everyday life. Anyone who has visited an art museum has had the experience of randomly wandering into an exhibition room and discovering an unfamiliar artistic style. From there, a greater interest may grow. Or, maybe it will recede and be forgotten. Regardless, there is a discovery freedom that commonly exists in physical life. The question is: How can this type of discovery be produced by algorithmic recommendations?

## 5. The Curiosity Engine

Recommendation system research in the area of discovery and creative freedom has been pursued under the title of curiosity (Ethics Workshop 2025), as well as serendipity (Fu *et al.* 2023). In both cases, what is sought are recommendations that are unfamiliar but also provocative and engaging.

Producing this curiosity is practically challenging, and also internally contradictory. The practical challenge derives from how typical recommendation algorithms are designed: they recycle established interests and preferences. The central strategy is "collaborative filtering," which means the recommenders screen for resemblances between options that a user has already enjoyed, and options that have not yet been selected. Then, once a similarity has been established, the resembling option is suggested (Munson *et al.* 2025).

The strategy of observing what a user likes and then finding more of the same makes intuitive sense. For someone enjoying a *Mission Impossible* film, the next offering may be another Tom Cruise action movie. For an aficionado of Dave Brubeck's jazz there may follow a Bill Evans track. In any case, what matters is that suggestions are built in a way that succeeds by *precluding* creative freedom. On the practical level, the entire process flows in the opposite direction, away from new possibilities and toward those already tried and proven. Inescapably, novel opportunities are filtered out as a byproduct of how recommenders are designed to function.

Besides being practically difficult, there is also a conceptual obstacle blocking recommendations geared for curiosity. It resembles the attempt to predict the unpredictable in the sense that recommendations must fall outside the scope of already established interests and behaviors, but they must nevertheless be inviting and attractive. They must be unexpected but also reassuring. Joining these qualities is challenging since we customarily find a proposal to be appealing *because* it resembles another that has already proven satisfying. However, if the aim is to generate dissimilar recommendations — options unlike previous satisfactions — then it is hard to know what elements of that unknown possibility will be agreeable, and which will not. The idea that a suggestion be both unfamiliar and engaging is almost a contradiction in terms.

To manage the problem and promote serendipity, some platforms pepper their recommendations with random selections (Munson *et al.* 2025). The hope is to intermittently present unfamiliar options that spur a new interest. Superficially, the strategy makes sense, but it also seems self–defeating: if a user is employing a recommender system, doesn't offering random possibilities undercut the entire idea? Isn't the original reason for recommenders precisely to escape haphazard selection?

One way to generate unfamiliar recommendations that are also engaging with a frequency that is higher than random has been explored under the title of the Curiosity Engine (Ethics Workshop 2025). And, one concept central to that work is antagonistic filtering (Brusseau 2025), which uses the same data and mechanisms of collaborative filtering, but adjusts the algorithmic priorities to privilege options that initially seem dissimilar to established preferences. The process starts from users divided into sets of divergent interests. Next, and across those separate groups, some few similarities are sought. Then, recommendations can be passed between them. These are suggestions that originate outside of established preferences, but that nevertheless promise some resonance with current tastes.

This resonance is not the comforting attraction of the familiar, but a response to provocation. When something is different but not entirely foreign, it can attract as a challenge, as a stimulation, as a subject of curiosity.

Because this is not a technical paper, the details of antagonistic filtering are secondary to the aim, which is to show that recommenders can promote user freedom by opening new horizons of interest.

## 6. Personal identity: Who we are and who we could be

Collaborative and antagonistic filtering move human experience in different directions. Going one way, recommenders are tuned for convenience, euphoria, and authenticity. They reinforce users' established preferences, and they create engagement through familiarity. Going the other way, recommenders are tuned to curiosity and creative freedom.

They function by generating interests that break away from who the user has been, and they produce engagement by being provocative.

This divergence is also a split in personal identity. It divides how identity is conceived. In one direction, the ideal individual life leads to an encounter with the true self, and then to the endeavor to remain consonant with that single identity. In the other direction, there is no true self, and the humanistic goal of recommender systems is to help users discover ways of becoming someone else. Instead of reinforcing what someone is, recommenders locate new identities to explore.

There is no indubitable way to select between these two human alternatives in today's digital reality. But, the contemporary default leans toward the first. The ethical and philosophical goal of establishing and stabilizing a person's single identity coheres well with the technical strategies that most recommendation systems currently employ. In accord with widespread collaborative filtering techniques, platforms seek their users' deep and true preferences, and then reinforce them by recommending similar offerings. Of course, the complexities of human personalities and corporate interests will blur this conclusion, but the foundational idea remains. The project is to identify and fortify a user's core inclinations.

Contrastingly, if human experience is turned toward discovery and the creation of new personal identities, then innovative work in recommendation algorithms will be required. Most important, research will need to proceed in the direction of provocative and serendipitous recommendations. The task will be to provoke user curiosity about new ways of being instead of reinforcing tastes and behaviors that define the identity that already exists.

Accompanying this technical research in the generation of curiosity, there is connected philosophical and ethical work to be done in the area of discontinuous personal identity. The theoretical question to be investigated is: How can selfhood be conceived as fundamentally disruptive? How can it be, in other words, that the basic condition is one of identities splitting away from their own pasts, with a multiplicity that is innate rather than being the result of imperfect or incomplete self–understanding? One source for guidance is David Hume's notion of selfhood as a bundle of interests, memories, and psychological states (Hume 1739). Another source is Gilles Deleuze's work on rhizomatic identities (Deleuze and Guattari 1980). A third direction is the idea of disintegration as the deepest source of selfhood (Brusseau 2019), and also the conception of genhumanism (Brusseau 2023).

Besides the technical work in recommendation algorithms and the philosophical work in fragmenting identity, there are pathways to open in law. Carlo Casonato (2023) has done introductory work in the development of a new human right: the right to discontinuity. The broad idea resembles earlier European stipulations in the area of privacy and the right to be forgotten. As currently established, the right to be forgotten is negative in the sense that it endows individuals with the ability to disappear from online reality and corporate data. Concretely, people can request the deletion of their personally identifying information from third–party databases. This disappearance — the forgetting — *is* the aim. By contrast, the right to discontinuity starts with the forgetting, but the aim is the ability to re–emerge online with new preferences and as a new collection of personally identifying information. So, the right to be forgotten is the idea that I can stop being who I am in digital reality, and the right to discontinuity is the idea that I can go on to become someone else.

Finally, dilemmas about the nature of human identity, about who we are and who we can be, are as old as the study of philosophy. What is different today, however, is that these questions are also choices that will be made for us — and done to us — by the AI technology of recommendation algorithms.

## Conclusion

Recommendation systems interact with human identity in two ways: they can find new possibilities within the spectrum of our expectations, and they can produce new spectrums of expectation. The first corresponds with recommenders as providing convenience and authenticity, while the second corresponds with recommenders as generators of curiosity and human freedom. Further, the dilemma is exclusionary, it is between the contentment of who we are at the cost of freedom, and the freedom to become someone else but at the cost of a convenience that is euphoric.

The results of this study are summarized in table 1.

| Human Experience of Recommendation Algorithms | |
|---|---|
| Euphoria Convenience | • The perfect choice offered at the critical moment: just before it is sought. There is no time for inconvenience, no time to ask what is wanted or how to get it.<br>• Euphoria convenience means the freedom to choose is literally an inconvenience. |
| Digital Authenticity | • Recommendation algorithms perfect self-knowledge from established behaviors. |
| Collaborative Filtering | • Generating recommendations and future preferences by recycling accurate understandings of past interests. |
| Antagonistic Filtering ("Creative Freedom") | • Unfamiliar but engaging recommendations.<br>• Recommendations that open ranges of potential recommending by provoking curiosity about new interests and experiences, as opposed to reinforcing established preferences and habits. |
| Personal Identity & Recommendation Algorithms | • Recommenders can generate new possibilities within the spectrum of current expectations, or they can produce new spectrums of expectation. The first corresponds with euphoric convenience and authenticity, the second with the human freedom of antagonistic filtering.<br>• The dilemma is between the contentment of being one person, and curiosity about becoming someone else. |

**Table 1:** The human experiences of recommendation algorithms.

## Bibliographic references

# DOBBIAMO *CREDERE* NELL'INTELLIGENZA ARTIFICIALE? UN DIALOGO PER COMPRENDERE IL PRESENTE CON MICHELA MILANO E PAOLO TRAVERSO

Paolo Costa, Eugenia Lancellotta e Boris Rähme

Abstract: This is the transcript of a round table discussion on the topic of trustworthy Artificial Intelligence (AI) organized and conducted by three FBK–ISR researchers involved in the project "Resilient Beliefs: Religion and Beyond." The experts consulted are Michela Milano and Paolo Traverso, two leading scholars in the field of Artificial Intelligence. Starting from the notion of resilient belief, topics such as the difference between human and artificial intelligence, the danger of replacing people with intelligent machines even in jobs traditionally considered creative, the geopolitical dimension of the AI Revolution, the legal regulation of AI hazards and the goal of consistent human oversight, how to create an epistemic and social environment conducive to the human–centric development of new digital technologies, are discussed.

Questa è la trascrizione di una tavola rotonda sul tema della *trustworthy AI* organizzata e condotta da tre ricercatori di FBK–ISR impegnati nel progetto "Resilient Beliefs: Religion and Beyond". Gli esperti consultati sono Michela Milano e Paolo Traverso, due studiosi di frontiera nel campo dell'Intelligenza Artificiale. Proprio a partire dalla nozione di convinzione resiliente vengono discussi temi quali: la differenza tra intelligenza umana e artificiale; il pericolo di sostituzione delle persone con macchine intelligenti anche in lavori tradizionalmente considerati creativi; la dimensione geopolitica della Rivoluzione dell'Intelligenza Artificiale; la regolamentazione giuridica dei rischi dell'IA e l'obiettivo di un sistematico *human oversight*; la creazione di un ambiente epistemico e sociale favorevole a uno sviluppo umanocentrico delle nuove tecnologie digitali.